\documentclass[times, twoside]{zHenriquesLab-StyleBioRxiv}
\usepackage[left]{lineno}
\leadauthor{Tanaka} 

\begin{document}

\hyphenpenalty=2000\relax
\exhyphenpenalty=2000\relax
\sloppy

\title{\LARGE Complex network prediction using deep learning}
\shorttitle{complex network prediction}

\author[1,2]{Yoshihisa Tanaka}
\author[3*\Letter]{Ryosuke Kojima}
\author[1]{Shoichi Ishida}
\author[1]{Fumiyoshi Yamashita}
\author[2,3*\Letter]{\\Yasushi Okuno}

\affil[1]{Graduate School of Pharmaceutical Sciences, Kyoto University}
\affil[2]{RIKEN Center for Computational Science, HPC and AI driven Drug Development Platform Division, Biomedical Computational Intelligence Unit}
\affil[3]{Graduate School of Medicine, Kyoto University}

\maketitle

\begin{abstract}
\section*{Abstract}
Systematic relations between multiple objects that occur in various fields can be represented as networks. Real-world networks typically exhibit complex topologies whose structural properties are key factors in characterizing and further exploring the networks themselves. Uncertainty, modelling procedures and measurement difficulties raise often insurmountable challenges in fully characterizing most of the known real-world networks; hence, the necessity to predict their unknown elements from the limited data currently available in order to estimate possible future relations and/or to unveil unmeasurable relations. In this work, we propose a deep learning approach to this problem based on Graph Convolutional Networks for predicting networks while preserving their original structural properties. The study reveals that this method can preserve scale-free and small-world properties of complex networks when predicting their unknown parts, a feature lacked by the up-to-date conventional methods. An external validation realized by testing the approach on biological networks confirms the results, initially obtained on artificial data. Moreover, this process provides new insights into the retainability of network structure properties in network prediction. We anticipate that our work could inspire similar approaches in other research fields as well, where unknown mechanisms behind complex systems need to be revealed by combining machine-based and experiment-based methods. 
\end {abstract}

\begin{keywords}
network prediction | scale-free | complex network | GCNs
\end{keywords}

\begin{corrauthor}
okuno.yasushi.4c@kyoto-u.ac.jp,\\kojima.ryosuke.8e@kyoto-u.ac.jp
\end{corrauthor}

\section*{Introduction}

\begin{figure*}[ht]
\begin{center}
\includegraphics[width=.9\linewidth]{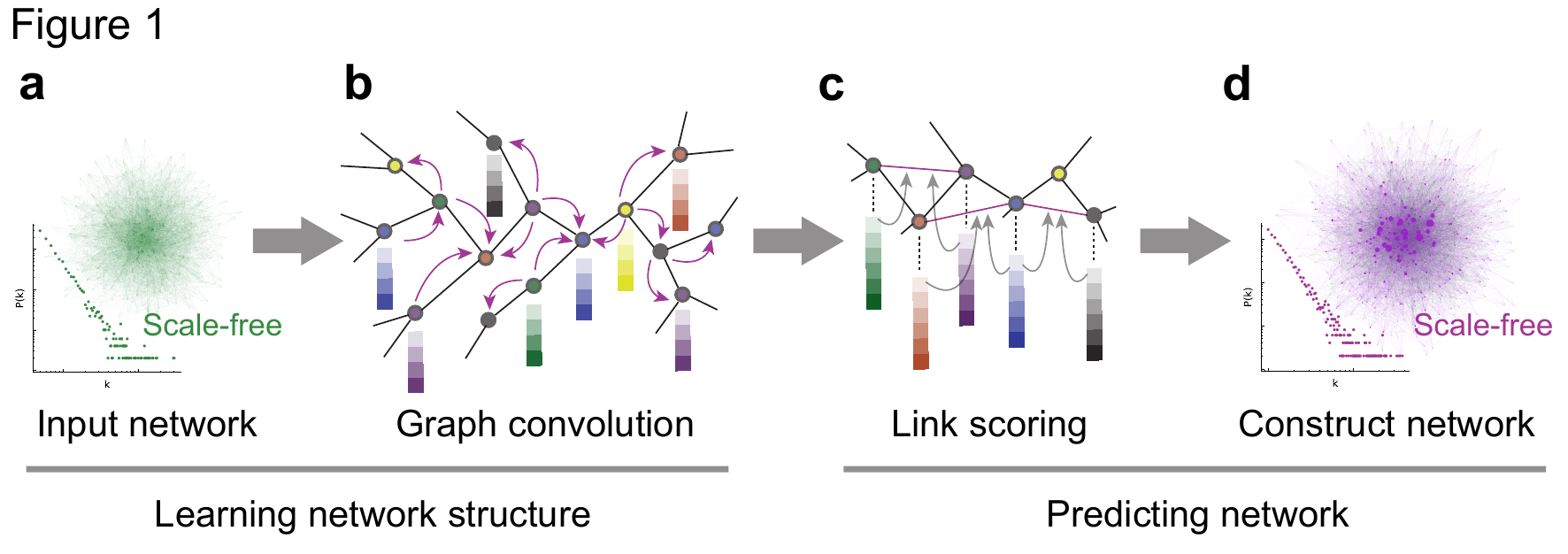}
\end{center}
\caption{\textbf{Illustration of the approach for network prediction.} (\textbf{a}) The network (green) represents an input network for training, (\textbf{b}) graph convolution part, the gradient blocks placed at individual nodes represent the image of feature vectors, (\textbf{c}) the representative predicted new links are represented as a highlighted edge (magenta) and (\textbf{d}) the predicted links are integrated into the input network to construct the predicted network (magenta).}
\label{fig1}
\end{figure*}

By network we understand a complex system with multiple relations between its components described as a graph, which plays a role in a variety of fields such as biology \cite{Watts1998-zy,Jeong2000-od,Fell2000-sk}, social sciences \cite{Amaral2000-we,Newman2001-rk,Aparicio2015-ab}, disease transmission \cite{Liljeros2001-im,Herrmann2020-xa} and the internet \cite{Albert1999-zv,Vazquez2002-ll}. Many real-world networks have a complex topology \cite{Albert2002-so}; however, there are laws that govern the structural property of the entire network, and therefore, understanding the network as a whole structure is important for elucidating the systems underlying it.

Two common topological traits of real-world networks are known as ‘small-world’  \cite{Watts1998-zy} and ‘scale-free’ \cite{Barabasi1999-nu}. The first of them means that the paths between any two nodes are short, and there are highly clustered connections \cite{Watts1998-zy} (e.g. your two friends are likely to be friends \cite{Backstrom2012-kz}). A network is scale-free when it has a few nodes with high connections and many nodes with low connections, where the degree distribution probability of nodes with $k$ connections, $P(k)$, follows a power law form as $p(k) \sim k^{-\gamma}$ ($\gamma$ is often between 2 and 4) \cite{Barabasi1999-nu}. Especially in the field of biology, the network topology was found to be indispensable for understanding the systematic behaviour of life \cite{Barabasi2004-zv}. In networks with these structural properties, highly connected nodes called ‘hubs’ plays a key role in fields when studying topics like network attack \cite{Albert2000-vh}, immunization strategy \cite{Pastor-Satorras2002-ew}, disease transmission \cite{Keeling2005-lq} and airline transportation \cite{Zanin2013-jk}, irrespective of biological domain \cite{Jeong2001-jj,Basso2005-tw,He2006-xk,Chen2008-gu}. Meanwhile, due to the uncertainty and difficulty in the measurement of real-world networks, it is still difficult to uncover true networks, and currently available networks are far from complete \cite{Kossinets2006-cr,Stumpf2008-pc,Clauset2008-vv}. Hence, there is a necessity for predicting whole networks from limited available data in order to estimate possible future relations or to complement unmeasured relations.

This prediction task is known as link prediction in the field of information retrieval \cite{Lu2011-my,Hamilton2017-ow}. Several studies have argued that the addition of the predicted links to the network should not change the original network topology \cite{Gomez2001-yz,Han2005-ld,Clauset2008-vv,Guimera2009-cs}; however, this conservative perspective is often overlooked in the pursuit of high accuracy in link prediction, which is one of the key aspects in the prediction of network structures as there is no guarantee that link prediction is equivalent to preserving properties of the overall network structure. For example, it is unclear whether just predicting individual links will lead to the retention of the structural features described above. 
 
In this study, we focus on preserving the structural property of networks and propose for the first time a deep learning approach using Graph Convolutional Networks (GCNs) to predict networks (Fig. 1) (see ‘Methods’). We show that although conventional link prediction methods could not predict the entire network while preserving its innate structure characteristics, GCNs do this due to their ability of learning network structures. These theoretical achievements are confirmed by applying our approach to biological scale-free and small-world real-world networks. Our results suggest that GCNs could assist to extrapolate a true network from current network data while maintaining its structure.

\section*{Results}

\subsection*{GCN reconstructed network topology of scale-free and small-world}

To examine whether the GCN-predicted network had similar structural properties to the original one, we firstly carried out network prediction on the artificially-generated networks. We used three model networks: the most classic Erdős-Rényi (ER) model (random network) exhibiting Poisson distribution \cite{Erdos1960-iq}, the Watts-Strogatz (WS) model (small-world network) \cite{Watts1998-zy} and the Barabási-Albert (BA) model (scale-free network) showing power law distribution \cite{Barabasi1999-nu}. These network models feature completely different patterns in the degree distribution (Fig. 2a), and their properties are summarized in Supplementary Table 1. To compare the GCN-based network prediction with the outcomes of other approaches, we additionally employed the DistMult- and Inner Product (IP)-based network prediction methods (see ‘Methods’) \cite{Yang2014-ki,Koren2008-iu}.

We found that, for the BA model network, the degree distribution of the GCN-predicted network exhibited power law form, different to the DistMult- and IP-predicted networks (Fig. 2a). As the number of links in the predicted network can be arbitrarily set, for fair comparison, we required the graph density of the predicted network to be almost similar to the one of the source network (Supplementary Table 1, 2). To see whether this tendency is consistent regardless of the number of links in the predicted network, we further investigated how the degree distribution of the predicted networks is affected as the number of links increases. While the DistMult- and IP-predicted networks gradually came near random graph-like degree distributions, the GCN-predicted networks consistently followed the power law (Supplementary Fig. 1-3). This suggests that GCNs preserve the scale-free property unlike DistMult and IP, probably due to the differences in network structure learning. 

Moreover, to evaluate the small-world effect for the predicted networks, we measured the clustering coefficient, an indicator of small-world property \cite{Watts1998-zy}. Among the three model networks, the WS model showed this property, with a high clustering coefficient \cite{Watts1998-zy} (Supplementary Table 2). We found that, regardless of model network, the clustering coefficient of the GCN-predicted networks elevated as the number of links increased, whereas that of the IP-predicted networks decreased (Supplementary Fig. 4). Moreover, the DistMult-predicted networks held low clustering coefficients. This showed that GCN-predicted networks consistently tend to exhibit small-world property compared to DistMult and IP-predicted ones (Supplementary Fig. 4). 

To confirm the validity and plausibility of the new links of the predicted network, we examined an enrichment curve. The enrichment analysis shows that while there is no difference in the outcomes of the three methods for the ER model network, GCN outperforms DistMult and IP for the BA and WS model networks (Fig. 2b-d). Thus, the internal validation of our approach for network prediction using the artificially-generated networks shows that for the BA model network, the GCN-predicted networks exhibit the scale-free and small-world properties, with high validity of the predicted links, indicating a possibility that GCNs are able to predict networks while preserving their original topology.

\begin{figure*}[ht]
\begin{center}
\includegraphics[width=.8\linewidth]{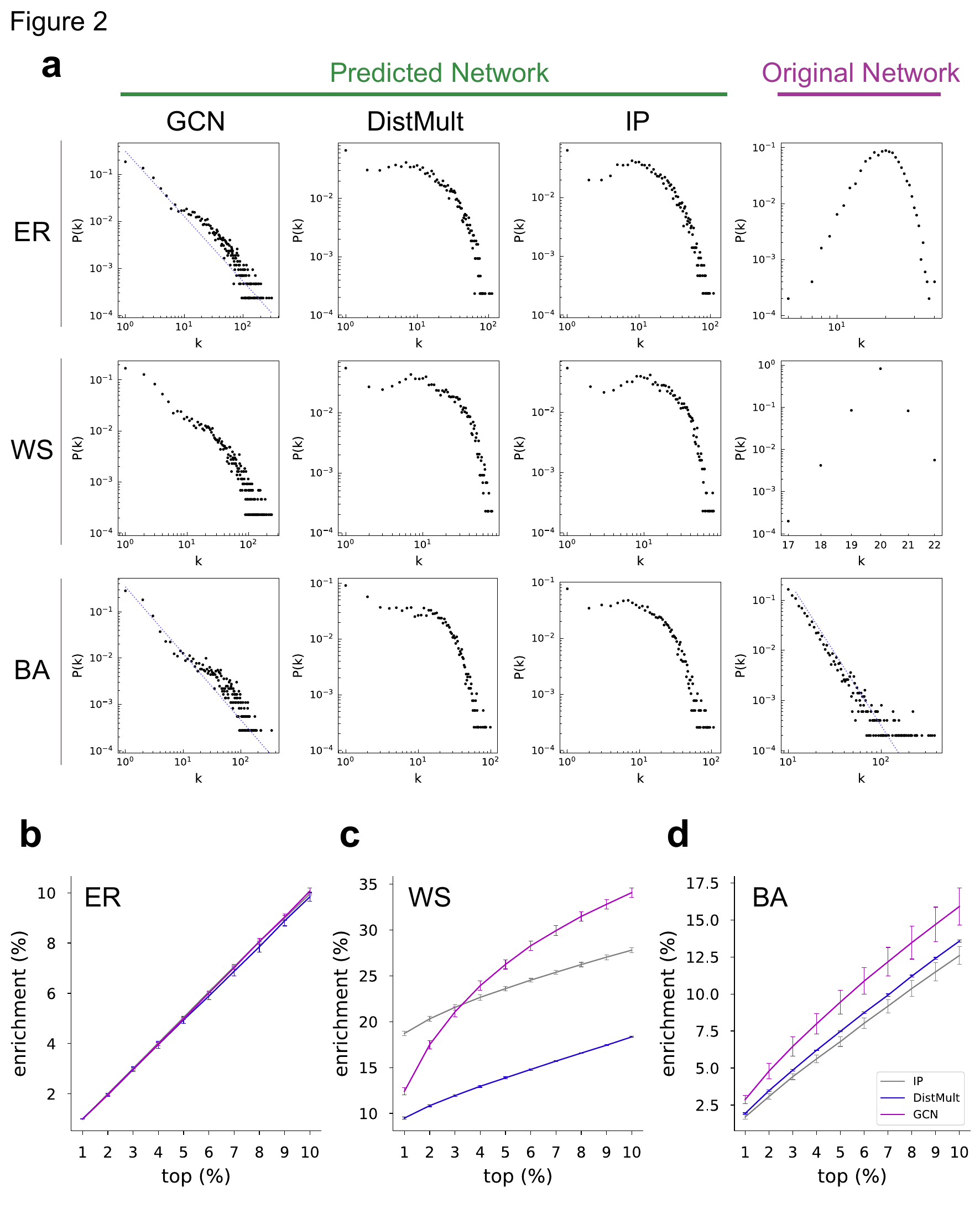}
\end{center}
\caption{\textbf{Evaluation of the predicted network for the three model networks.} (\textbf{a}) The degree distributions of the predicted networks for the three model networks (BA/WS/ER) with the three methods (GCN/DistMult/IP) (green labelled columns). The panels with dotted lines (blue) show the degree distribution following the power law. The link threshold was adopted as 40000 for ER, 40000 for WS, 30000 for BA. The original degree distributions of the artificially-generated model networks were displayed at the original network panel (magenta labelled columns) and their network properties were listed in Supplementary Table 1. The original networks were comprised of edges (49900 for BA, 50000 for WS, 50160 for ER) and nodes (50000 for each network). The X axis represents the degree k, and the Y axis represents the degree distribution probability P(k). The detailed network properties of the predicted networks were listed in Supplementary Table 2. (\textbf{b, c, d}) The enrichment analysis with the three methods; IP (grey), DistMult (blue), GCN (magenta). The X axis represents an arbitrary link threshold. Error bar: mean ± standard deviation (n = 3).}
\label{fig2}
\end{figure*}

\subsection*{Network predictions of real-world biomolecular networks}

To verify the validity of our observations on real-world networks with more complex topological properties, we constructed six human biomolecular networks—‘expression’, ‘interaction’, ‘phosphorylation’, ‘state change’, ‘complex’ and ‘catalysis’—using a biological interactome data source \cite{Cerami2011-ie,Rodchenkov2020-yw}, and analysed their network properties (Supplementary Table 1). Four of the six networks exhibited the scale-free property (Fig. 3). Since their clustering coefficients and average shortest path lengths are relatively small compared to the graph diameters, each network is also presumed to have the small-world property (Supplementary Table 1). After carrying out network predictions for each biomolecular network, we noted that while the degree distribution of the GCN-predicted networks followed power law form for all the biomolecular networks, most of the DistMult- and IP-predicted networks did not (Fig. 3, Supplementary Fig. 5-10). Notably, the two networks that originally did not have the scale-free property (‘complex’ and ‘catalysis’) were also predicted as scale-free, which suggests that GCNs may have the potential to reconstruct scale-free characteristics that true biological networks should have. Similarly, the clustering coefficient analysis shows that all six GCN-predicted networks exhibit the small-world property (Supplementary Fig. 11). The enrichment analysis showed that GCNs outperform DistMult and IP for all the biomolecular networks, further evidence that the GCN-predicted networks are likely to be more plausible (Fig. 4a-f). Hence, the internal validation using real-world biological networks suggest that GCNs can reconstruct network while preserving their properties.

\begin{figure*}[p]
\begin{center}
\includegraphics[width=.8\linewidth]{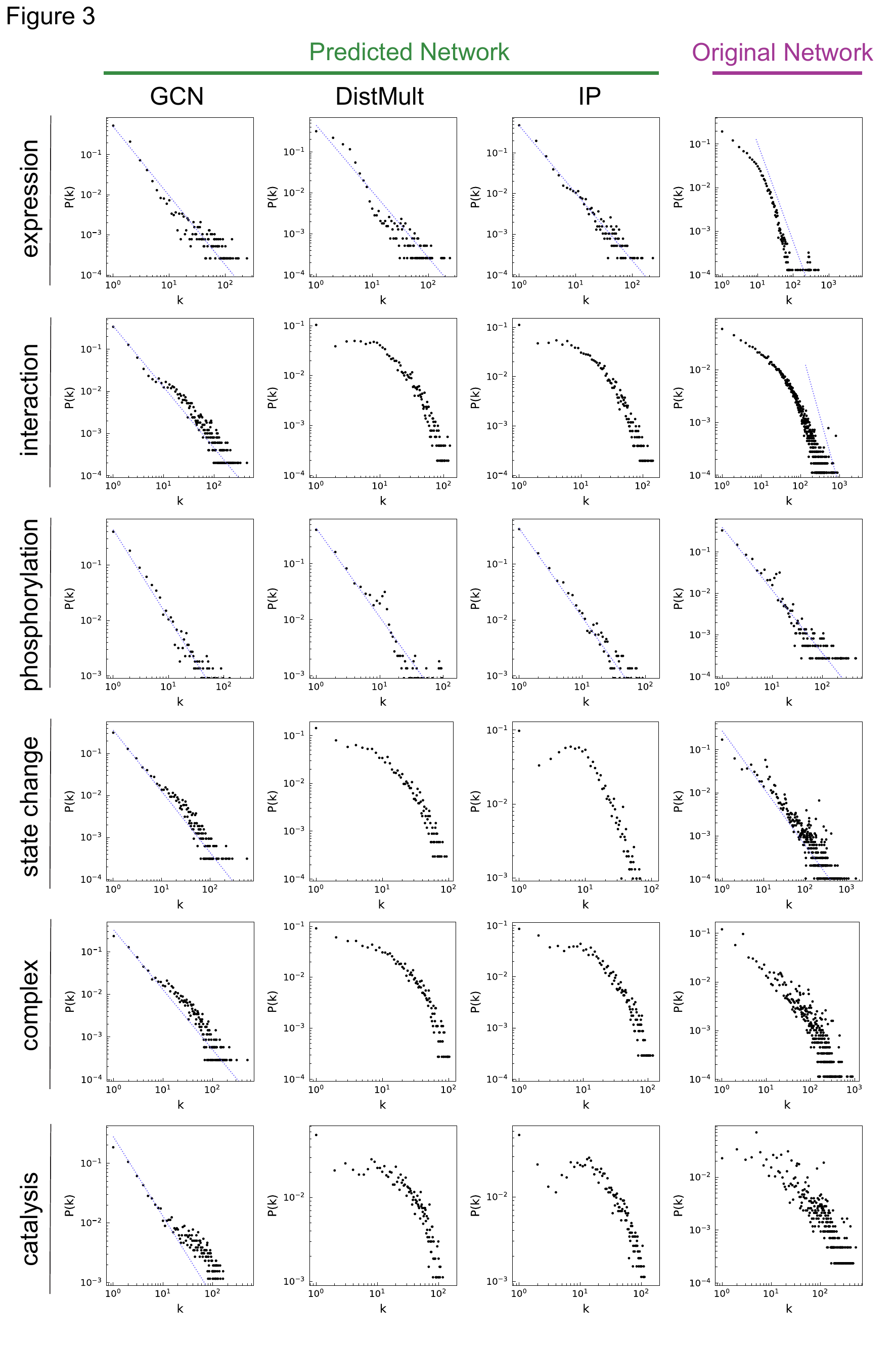}
\end{center}
\caption{\textbf{The degree distributions of the predicted networks for the six biomolecular networks} (green labelled columns). The link threshold was adopted as 10000 for expression, 40000 for interaction, 7000 for phosphorylation, 20000 for state change, 30000 for complex, 40000 for catalysis. The original degree distributions of the individual biomolecular networks were displayed at the original network panel (magenta labelled columns), and their network properties were listed in Supplementary Table 1. The X axis represents the degree k, and the Y axis the degree distribution probability P(k). The detailed network properties of the predicted networks were listed in Supplementary Table 3.}
\label{fig3}
\end{figure*}

\subsection*{External validation of the GCN-predicted networks using the other real-world networks}

\begin{figure*}[t]
\begin{center}
\includegraphics[width=.8\linewidth]{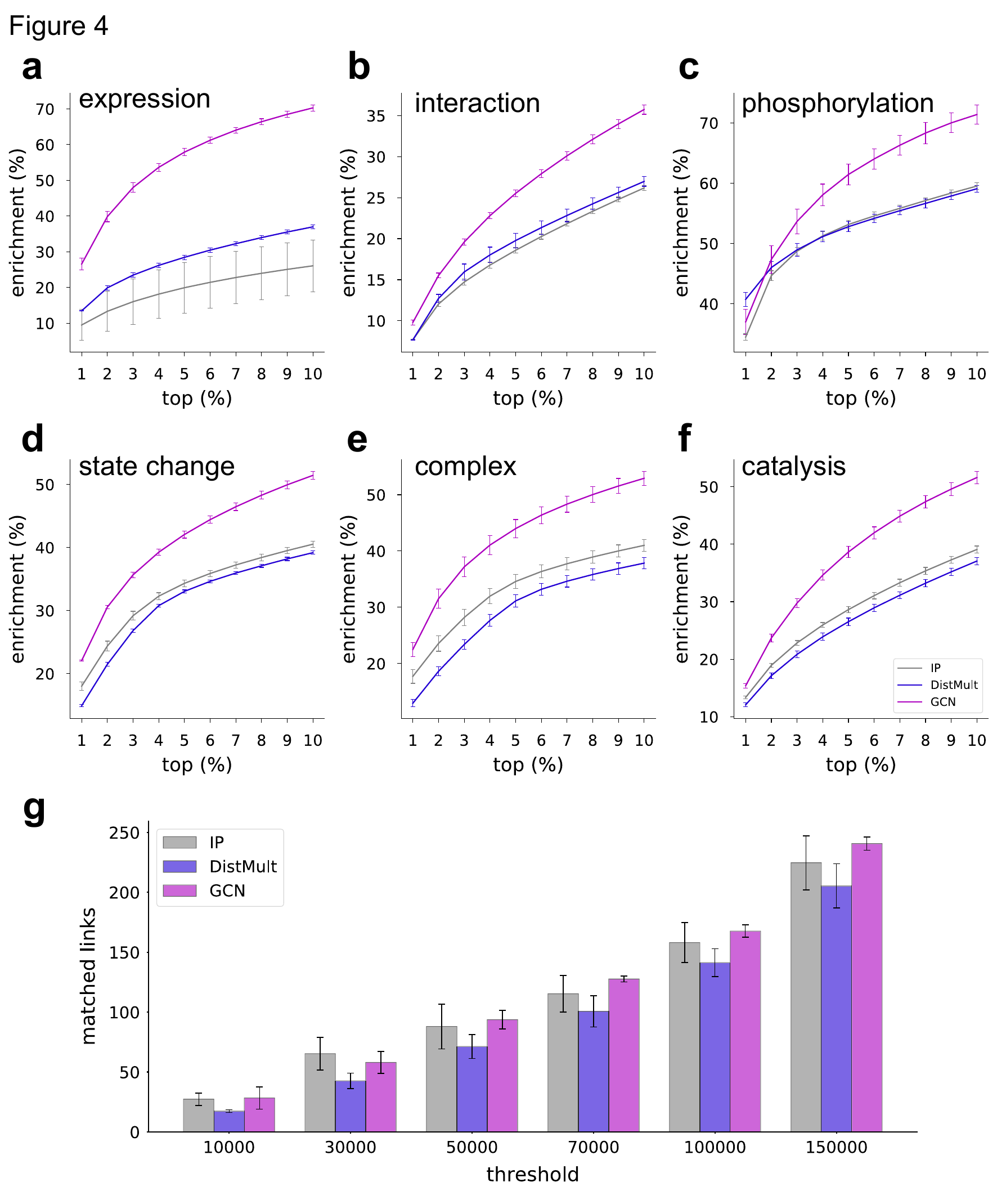}
\end{center}
\caption{\textbf{Performance evaluation of the predicted network for the six biomolecular networks.} (\textbf{a-f}) The enrichment analysis with the three methods: IP (grey), DistMult (blue), GCN (magenta). The X axis represents an arbitrary link threshold. (\textbf{g}) The comparison of matching links of the predicted ‘interaction’ biomolecular network with the HuRI network by the three methods: IP (grey), DistMult (blue), GCN (magenta).The X axis represents the link threshold for the predicted networks, while the Y axis represents the number of the matched links. Error bar: mean ± standard deviation (n = 3).
}
\label{fig4}
\end{figure*}

To examine the true validity of the predicted network, we conducted an external validation, too, by comparing the predicted network above using the real-world biological network with another external interactome network obtained from experiments. We investigated the latest experimental interactome network of the human reference interactome (HuRI) \cite{Luck2020-hb}, noting its scale-free and small-world characteristics (Supplementary Fig. 12a, Supplementary Table 1). Among the six considered biomolecular networks, the ‘interaction’ one is most similar to HuRI (Supplementary Fig. 12b). The analysis of the extent to which the predicted network generated from ‘interaction’ (Fig. 3, Supplementary Fig. 6) matches HuRI shows that while DistMult is inferior to IP and GCNs, the degree of matching of the IP- and GCN-predicted networks seems to be almost comparable (Fig. 4g). However, given that the IP-predicted networks do not show scale-free on the structural level (Fig. 3, Supplementary Fig. 6), even if their degree of matching is better to some extent (Fig. 4g), they do not reflect the intrinsic structure of the HuRI network. In contrast, the GCN-predicted network satisfies both the high degree of matching and structural homology with the HuRI network (Fig. 3, Fig. 4g, Supplementary Fig. 6). This evidence further supports that network prediction by GCNs is more effective for predicting real-world networks than its currently known counterparts.




\section*{Discussion}

We propose a deep learning approach for network prediction (Fig. 1) based on GCNs that predicts complex networks while preserving their structural properties (Fig. 2, Fig. 3) Experiments with biological networks guarantee the robust validity of this new method, also showing that it outperforms its currently standard counterparts (Fig. 4). As most real-world networks exhibit complex structural properties \cite{Albert2002-so}, we focused in our study on the importance of their retention in order to capture the hidden mechanisms behind them. Our results indicate that the deep learning-based GCNs are able to capture the structural features of the original networks by learning their structural information. Previous work demonstrated that it is difficult to infer the scale-free property of a network by partial network sampling due to the small coverage of the protein–protein interaction networks \cite{Han2005-ld}. Our analysis reveals that the emerging deep learning techniques can contribute to overcome the challenge of predicting networks while preserving their topology. Moreover, we also expect the graph embedding used in GCNs to be useful in machine learning tasks, such as graph or node classification \cite{Hamilton2017-ow}. In the process of expanding the applicability range of the graph embedding, GCN could prove to be useful in achieving better graph representations by learning the network structure information \cite{Kipf2016-fg,Schlichtkrull2017-si}, as suggested by our study.

However, the limitations of our approach in predicting networks essentially lie in the training dataset. New external nodes cannot be added to the predicted network and consequently it is almost infeasible to predict nodes not included in the current data that could be discovered in the future. The scale-free network model is also known as the preferential attachment model \cite{Barabasi1999-nu} in which new nodes are preferentially attached to highly connected ones forming nodes with higher degree, which eventually grows the network. Although GCNs might potentially learn these processes, this point of handling new nodes is slightly different from the BA model. By extending the framework to address this issue, network prediction would be more in line with the nature of scale-free networks. Our work sheds light on a landscape for investigating unknown mechanisms behind complex systems by combining machine-based and experiment-based methods.

\section*{Methods}

\subsection*{Development of an approach for network prediction}

The proposed approach for network prediction is presented in Fig. 1. The key idea was inspired by graph embeddings, where graph-structured data is projected onto vector spaces. A node receives a low-dimensional vector representation to reflect the relational data through a process known as representation learning \cite{Hamilton2017-ow}.

 While different embedding approaches have been studied in link prediction \cite{Bordes2013-rt,Yang2014-ki,Nickel2016-kj,Trouillon2016-hx,Nickel2017-jk}, the remarkable progress of artificial intelligence (AI) research in recent years, in particular of deep learning \cite{Kipf2016-fg,Velickovic2017-fb,Xu2018-rj,Xu2018-zy}, raises hopes towards more accurate predictions. In our investigation, we employ to this end GCNs \cite{Kipf2016-fg} due to their performance in various prediction tasks \cite{Schlichtkrull2017-si}. 

The principle behind GCNs is conceptually similar to the one of Convolutional Neural Networks (CNNs), which brought a breakthrough in imaging. A CNN learns image features by convolving surrounding pixel information for a pixel \cite{Krizhevsky2012-lm}, whereas a GCN learns graph topological features by convolving adjacent node information of a node \cite{Kipf2016-fg}. Inspired by CNNs’ ability to restore original images \cite{Masci2011-np}, we presumed that GCNs can reconstruct network topological characteristics by learning the network structure. To test this hypothesis, we implemented a GCN and developed an approach for network prediction, which is explained step-by-step below and illustrated in Fig. 1. The input is a network data with its characteristic structure (Fig. 1a). The topological structure of the network is learned by the GCN, and this topological information is transformed into a vector space, the individual nodes of which are represented as feature vectors (Fig. 1b). Next, by using these feature vectors, link scores are calculated for unknown links (Fig. 1c). As the links with high scores are more likely to appear in the future, they are extracted in a descending order and integrated into the input network, constructing thus the predicted network (Fig. 1d). We defined a set of these sequential steps (starting with the network as input and ending with the predicted network as output) as {\it network prediction}.

\subsection*{Learning architecture}

The core learning frame of network prediction followed an already established path \cite{Schlichtkrull2017-si}. The learning architecture mainly consists of two parts, graph embedding as encoder and link scoring as decoder. In the process of graph embedding (Fig. 1b), the graph convolution technique was used for the GCN to encode an input network as a vector. Our implementation followed Kipf’s model \cite{Kipf2016-fg} with two graph convolution layers and the activation function ReLU. In the process of link scoring (Fig. 1c), the encoded vector data is decoded by the scoring function as a score that represents how likely an unknown link is to exist. For the GCN and IP models, the score of a possible link is computed as a dot product of the feature vectors of the two nodes at a given link \cite{Koren2008-iu,Kipf2016-lu}. For the DistMult model, the weighted score is computed following the matrix factorization algorithm, which is a standard benchmark for link prediction \cite{Yang2014-ki}. To allow nodes acquire more accurate representations, the model is trained to gain a larger score for an existing link (positive link) than for a non-existing one (negative link). As it is impossible to know whether non-existing links in a current network will appear in the future or not, negative links were randomly sampled as previously established \cite{Yang2014-ki,Trouillon2016-hx,Schlichtkrull2017-si}. Since we assumed that the quantity of available data is limited and small \cite{Kossinets2006-cr,Stumpf2008-pc}, the input network data were divided into 5000 links for training and the remaining for testing. In the process of constructing a predicted network (Fig. 1d), the links with high scores were integrated into the training data in a descending order. The predicted network is determined as a network consisting of this set of links. The number of links to be assigned to the predicted network can be a priori set as a link threshold. The procedure of network prediction was independently performed three times with different datasets. The learning parameters are listed in Supplementary Table 4. We implemented the method in python, integrating it into our open source GCN platform, kGCN \cite{Kojima2020-up}.

\subsection*{Enrichment analysis}

Enrichment was defined as the ratio between {\it the number of test links in top N links} and {\it the total number of links}, where $N$ is the arbitrary link threshold, and the total number of links is $n(n-1)/2$, $n$ being the number of total nodes in the input network. This is an indicator of the precision performance of link prediction. The higher the value, the more precisely the links are predicted.

\subsection*{Network dataset}

To test the proposed approach we consider 10 networks, both artificially generated and real world. Three model networks (BA model, WS model and ER model) were generated using the python package {\it networkx} \cite{Hagberg2008-vu} in order to share approximately the same graph density. We also consider some biomolecular networks whose original dataset (version 11) was downloaded from Pathway Commons \cite{Cerami2011-ie,Rodchenkov2020-yw}. We selected the six types of graph datasets: ‘control expression of’, ‘interact with’, ‘control phosphorylation of’, ‘control state change of’, ‘in complex with’, ‘catalysis precedes’. Each dataset was preprocessed by conversion into an undirected graph and removal of selfloops. The HuRI network was extracted from The Human Reference Protein Interactome Mapping Project \cite{Luck2020-hb}. Since the names of nodes were written in Ensembl gene identifier, these were converted into gene symbols to correspond with the descriptor of Pathway Commons using the python package {\it mygene} \cite{Xin2016-ml}, after which the selfloops were removed. All the networks were represented as undirected and simple graphs, and the detailed network properties are summarized in Supplementary Table 1.

\subsection*{Network property analysis}

The power law fitting in degree distribution was performed using the python package {\it powerlaw} \cite{Alstott2014-rj}. To confirm the fitting state of power law, the exponential function was employed for the fitting comparison. The power law fitting was determined when the likelihood value was positive. Network properties \cite{Costa2007-ws} (clustering coefficient, average shortest path, degree assortativity coefficient, diameter, graph density, average degree, number of components) were calculated using python package {\it networkx} \cite{Hagberg2008-vu}. When multiple connected components were observed in a network, the network property analysis was performed on the largest one. Network visualization was performed using {\it Cytoscape} \cite{Shannon2003-va}.

\subsection*{Data availability}

The public network datasets used in this study are freely downloaded at Pathway Commons (\href{https://www.pathwaycommons.org/}{https://www.pathwaycommons.org/}) and The Human Reference Protein Interactome Mapping Project (\href{http://www.interactome-atlas.org/}{http://www.interactome-atlas.org/}). The three model networks generated in this study are included in the GitHub repository.

\subsection*{Code availability}

The whole code for network prediction is available at our open source GCN platform for lifescience, kGCN (\href{https://github.com/clinfo/kGCN}{https://github.com/clinfo/kGCN}).

\begin{acknowledgements}
We thank Y. Tamada for helpful discussions. This study was supported by RIKEN Junior Research Associate Program; Medical Science Innovation Hub Program of RIKEN; AMED under Grant Number P20kk0205013; Cabinet Office, Government of Japan, Public/Private R$\&$D Investment Strategic Expansion Program (PRISM). This paper format was generated through self-modification of the original template designed by Ricardo Henriques.
\end{acknowledgements}

\begin{contributions}
Y.T., R.K. and Y.O. designed the experiments. Y.T. and R.K. conducted the experiments. Y.T., R.K. and S.I. wrote the codes. Y.T., R.K., S.I. and Y.O. analysed the results. Y.T. and Y.O. wrote the original manuscript. R.K., S.I. and F.Y. reviewed and edited the manuscript. F.Y. and Y.O. supervised the study.
\end{contributions}

\begin{interests}
The authors declare no competing or financial interests.
\end{interests}

\section*{Bibliography}
\bibliography{NetPred_references.bib}


\captionsetup*{format=largeformat}

\renewcommand{\figurename}{Supplementary Fig.}
\setcounter{figure}{0}

\makeatletter
\renewcommand{\thefigure}{\@arabic\c@figure}

\begin{figure*}[p]
\begin{center}
\includegraphics[width=.8\linewidth]{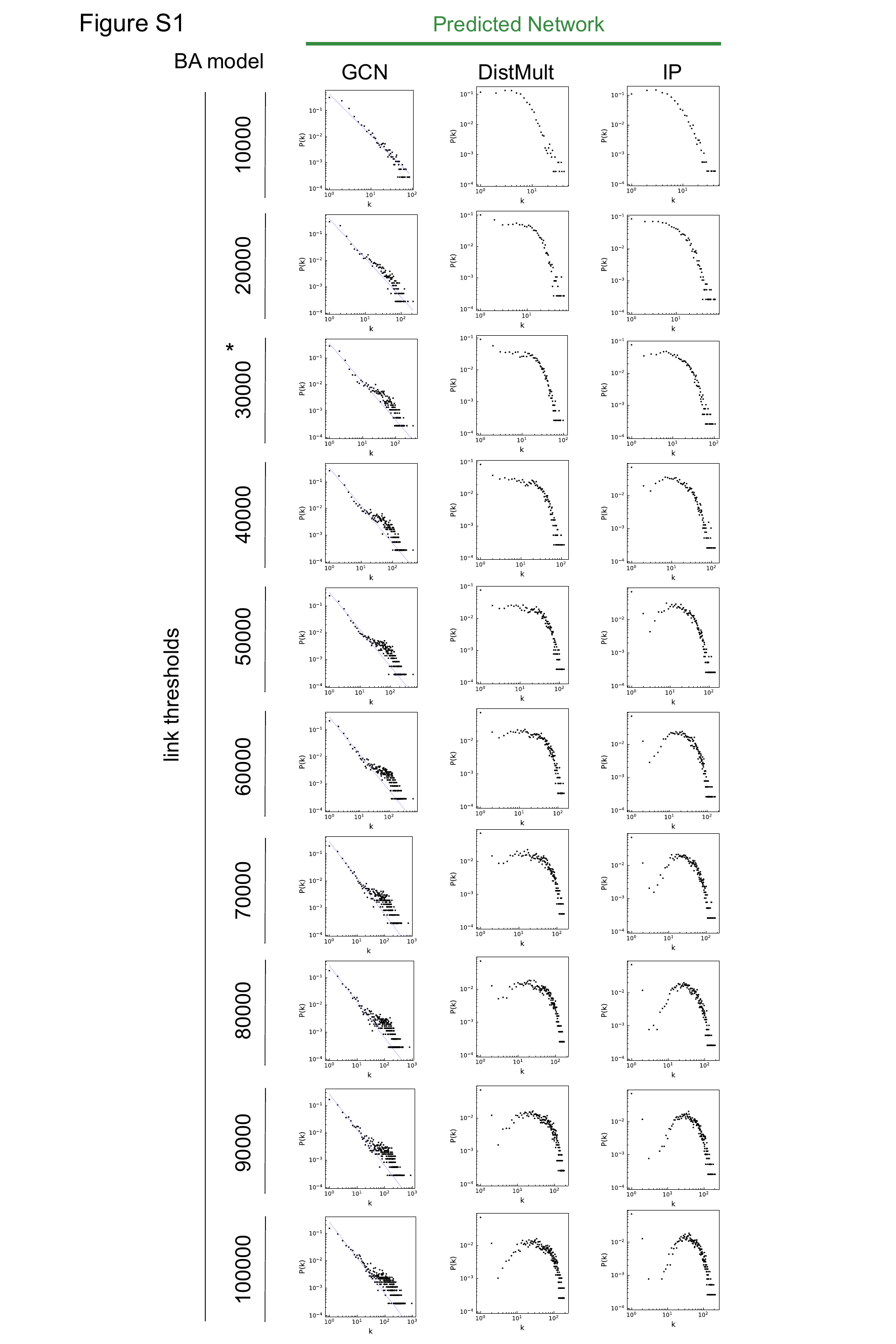}
\end{center}
\caption{\textbf{The degree distributions of the predicted networks with the sequential link thresholds for the BA model network.} The panels with dotted lines (blue) signal that the degree distribution follows the power law. The marked panels were used in Fig. 2. The detailed network properties of the predicted networks were listed in Supplementary Table S2.
}
\label{figS1}
\end{figure*}

\begin{figure*}[p]
\begin{center}
\includegraphics[width=.8\linewidth]{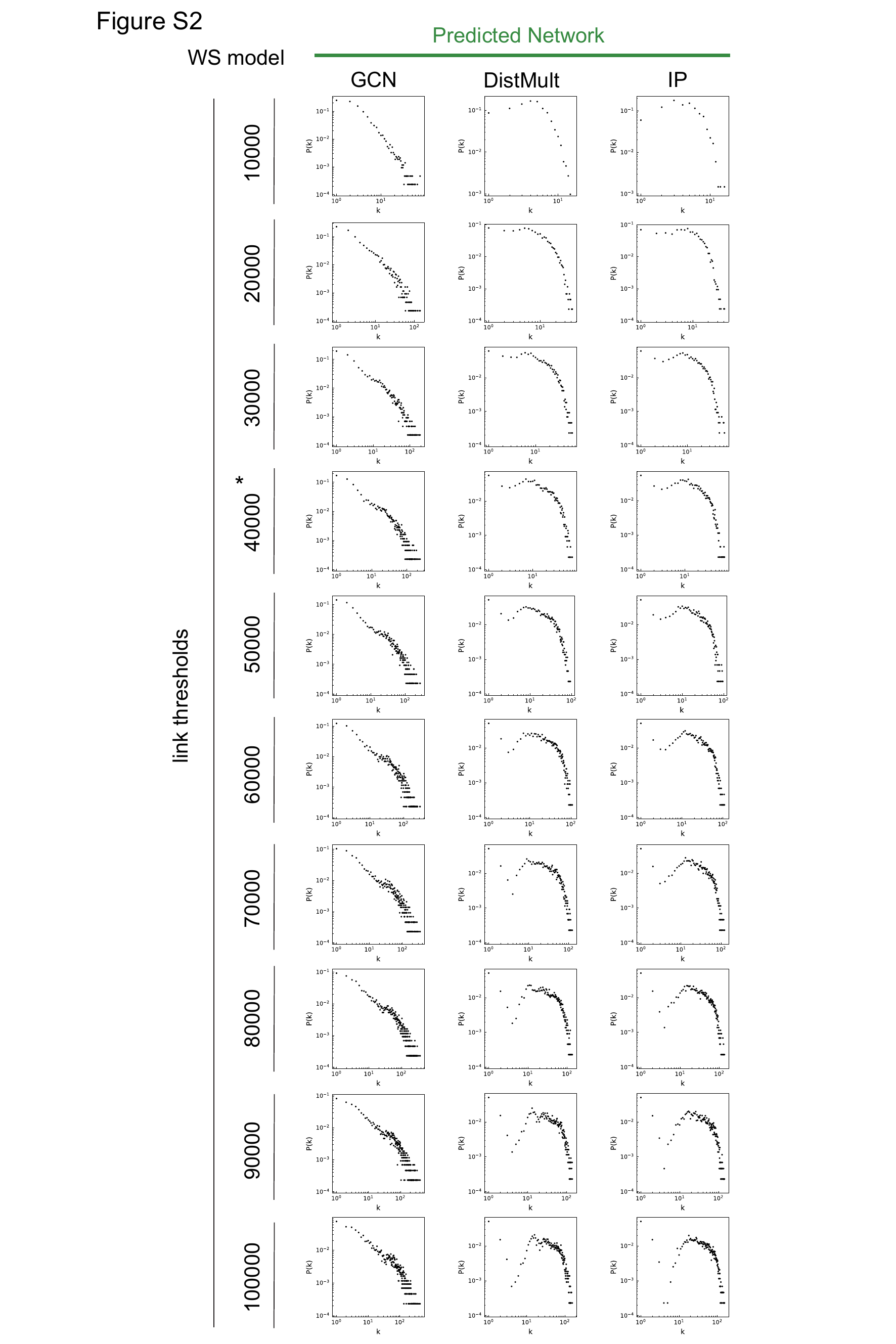}
\end{center}
\caption{\textbf{The degree distributions of the predicted networks with the sequential link thresholds for the WS model network.} The panels with dotted lines (blue) signal that the degree distribution follows the power law. The marked panels were used in Fig. 2. The detailed network properties of the predicted networks is listed in Supplementary Table S2.
}
\label{figS2}
\end{figure*}

\begin{figure*}[p]
\begin{center}
\includegraphics[width=.8\linewidth]{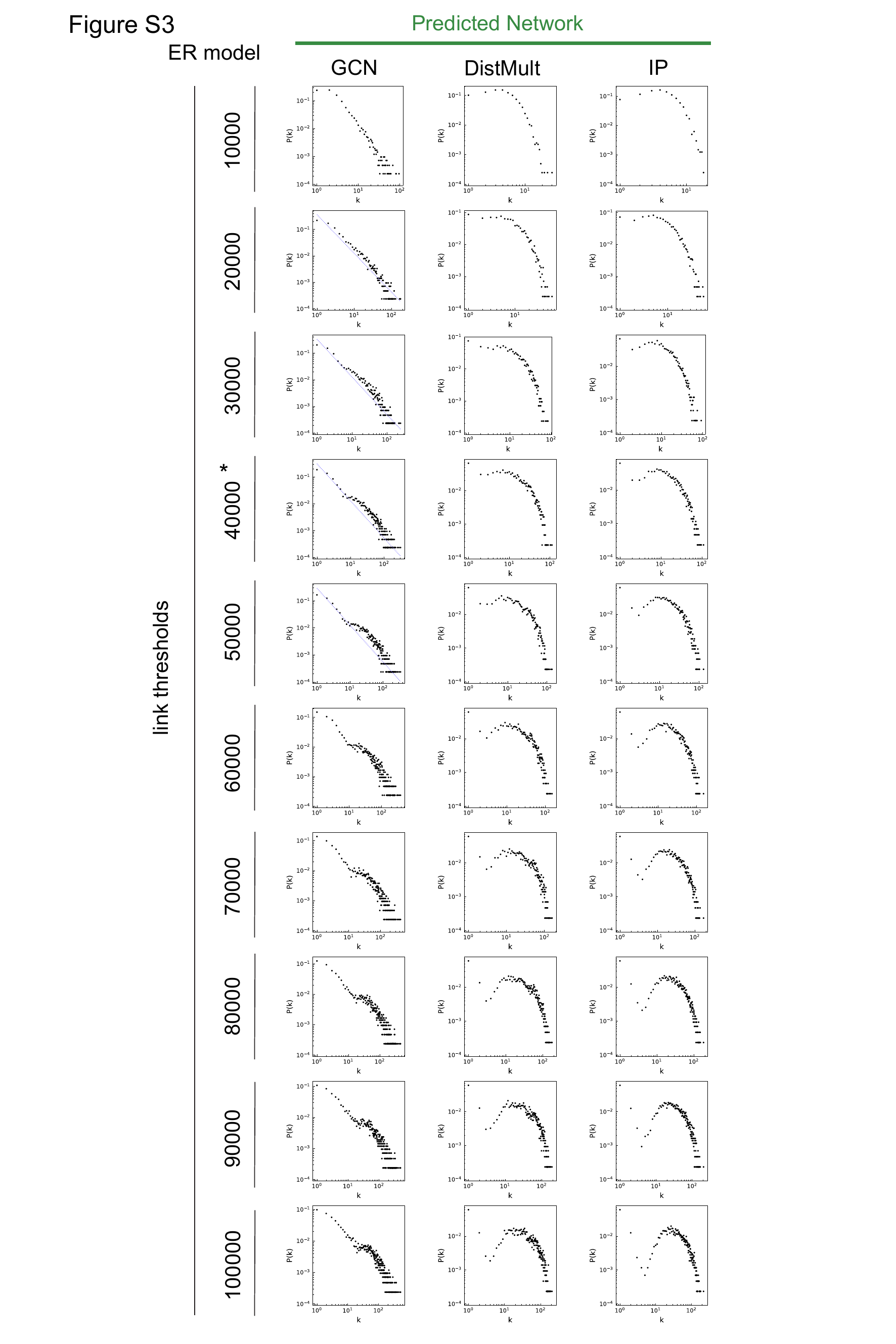}
\end{center}
\caption{\textbf{The degree distributions of the predicted networks with the sequential link thresholds for the ER model network.} The panels with dotted lines (blue) signal that the degree distribution follows the power law. The marked panels were used in Fig. 2. The detailed network properties of the predicted networks is listed in Supplementary Table S2.
}
\label{figS3}
\end{figure*}

\begin{figure*}[p]
\begin{center}
\includegraphics[width=\linewidth]{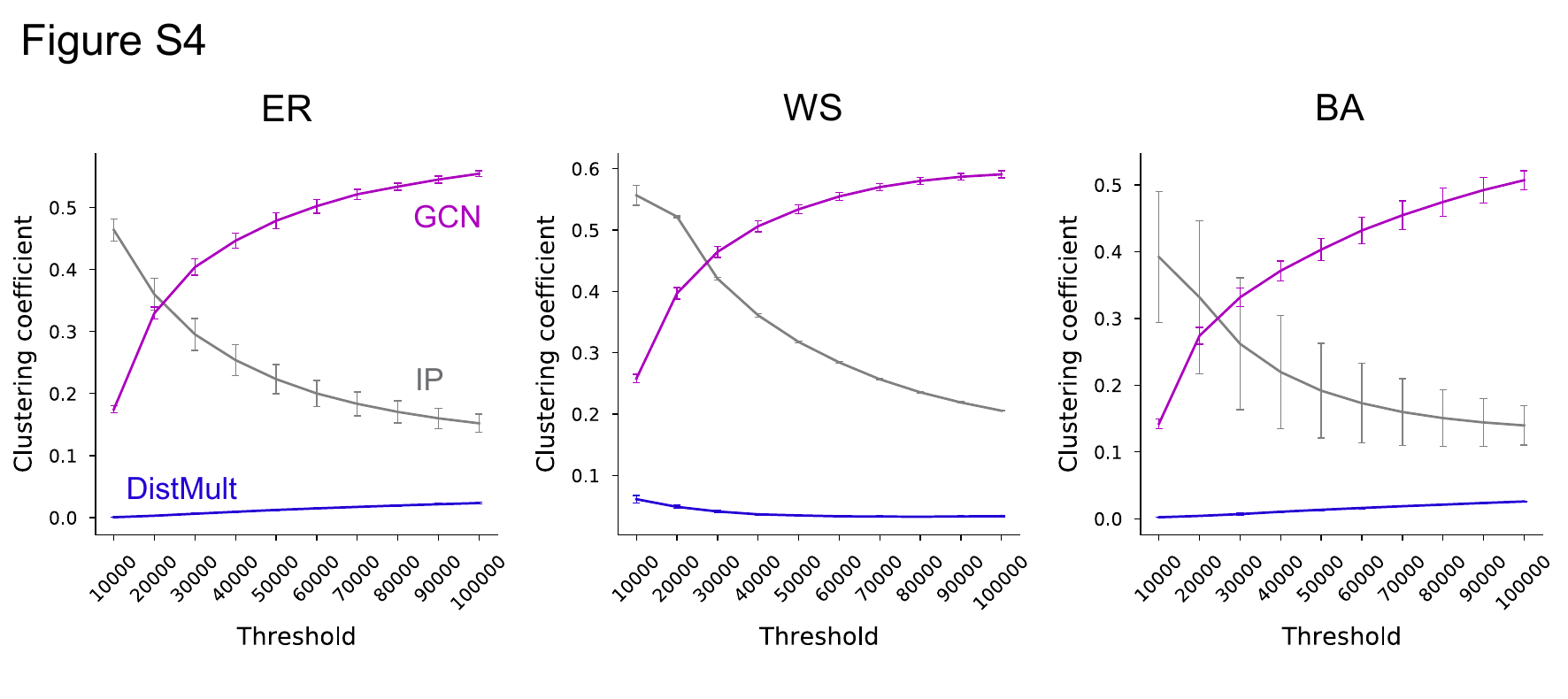}
\end{center}
\caption{\textbf{The transition of clustering coefficient for the three model networks.} The X axis represents the link threshold for the predicted networks. The detailed value is listed in Supplementary Table 2. Error bar: mean ± standard deviation (n = 3).
}
\label{figS4}
\end{figure*}

\begin{figure*}[p]
\begin{center}
\includegraphics[width=.9\linewidth]{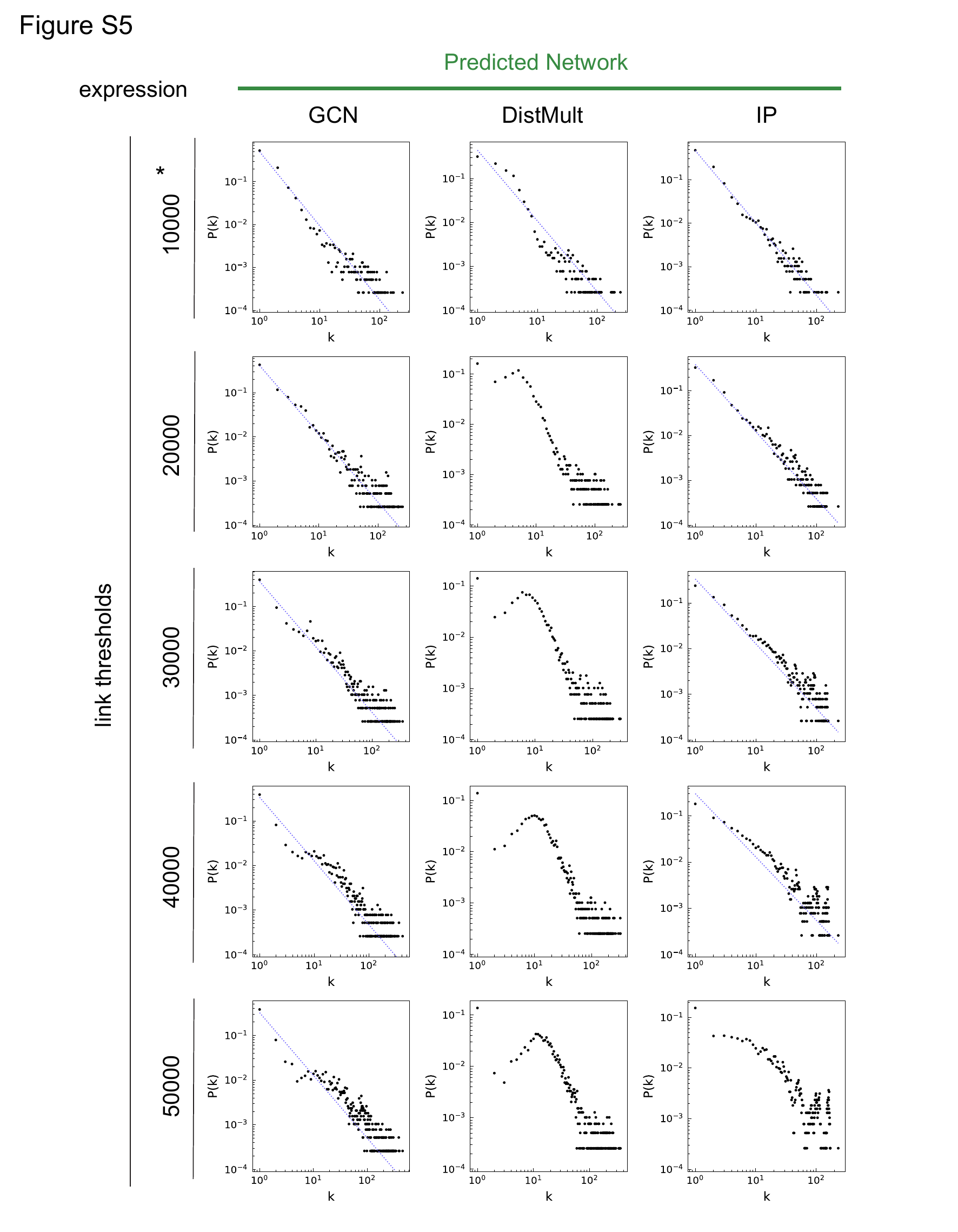}
\end{center}
\caption{\textbf{The degree distributions of the predicted networks with the sequential link thresholds for the ‘expression’ biomolecular network.} The panels with dotted lines (blue) signal that the degree distribution follows the power law. The marked panels were used in Fig. 3. The detailed network property is listed in Supplementary Table 3.
}
\label{figS5}
\end{figure*}

\begin{figure*}[p]
\begin{center}
\includegraphics[width=.9\linewidth]{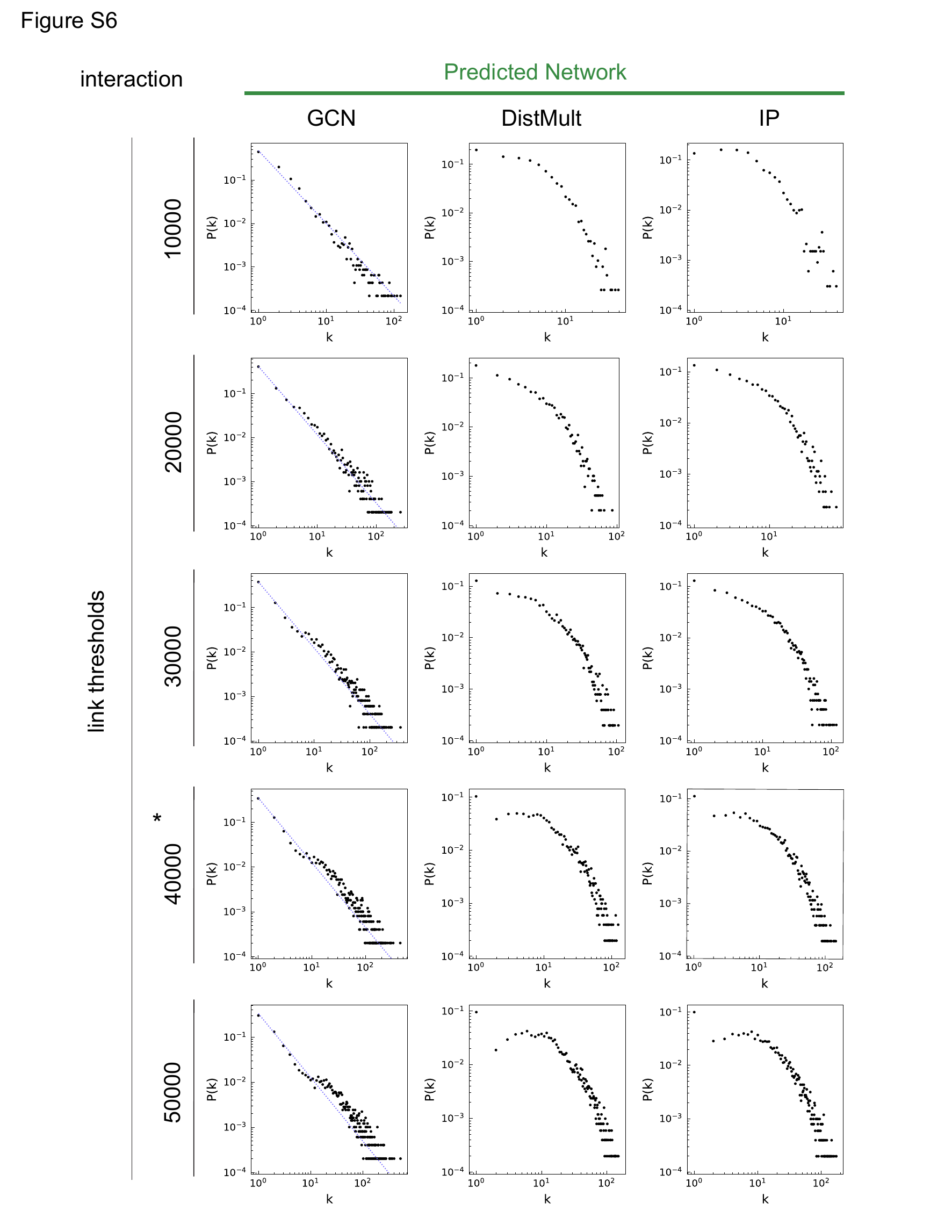}
\end{center}
\caption{\textbf{The degree distributions of the predicted networks with the sequential link thresholds for the ‘interaction’ biomolecular network.} The panels with dotted lines (blue) signal that the degree distribution follows the power law. The marked panels were used in Fig. 3. The detailed network property is listed in Supplementary Table 3.
}
\label{figS6}
\end{figure*}

\begin{figure*}[p]
\begin{center}
\includegraphics[width=.9\linewidth]{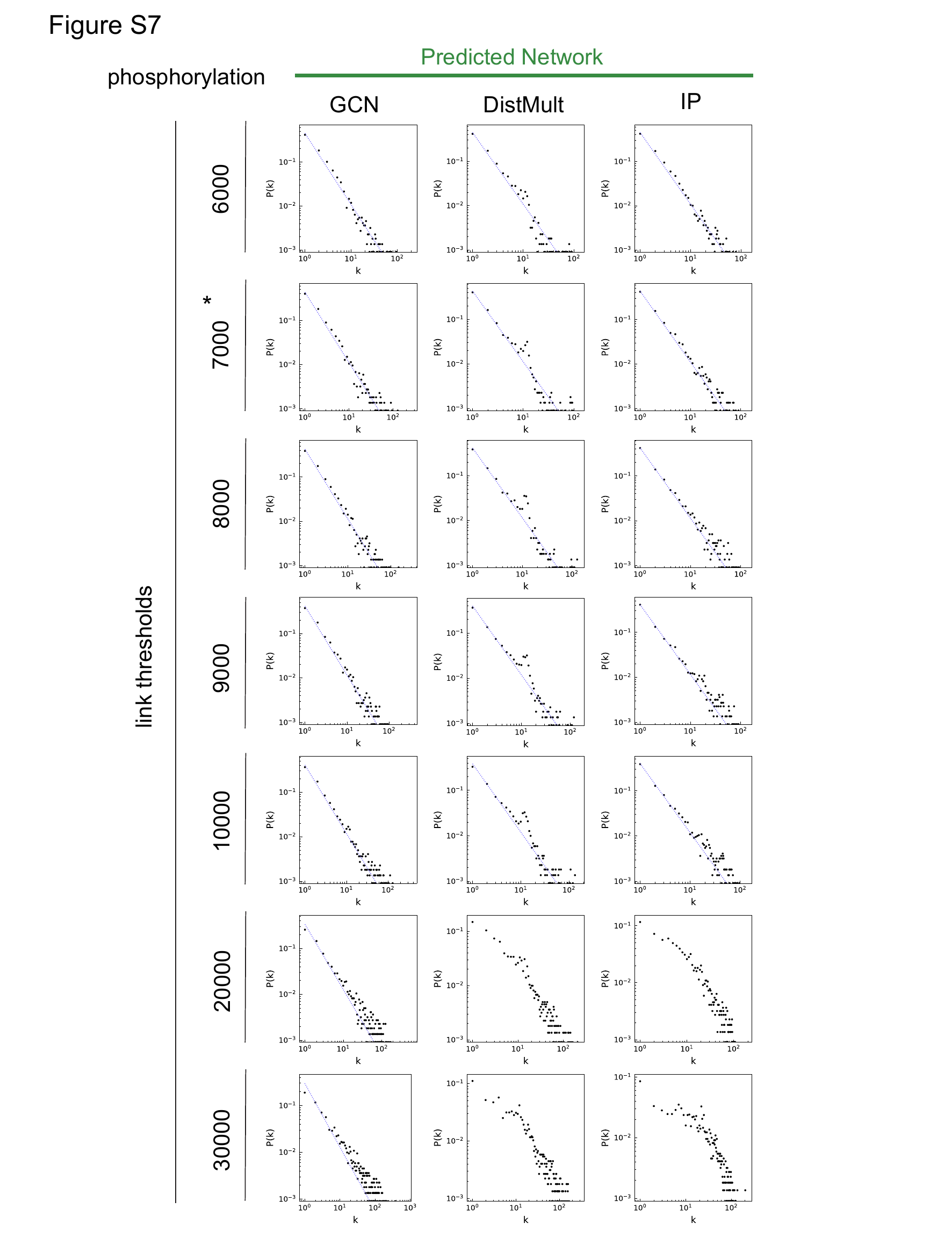}
\end{center}
\caption{\textbf{The degree distributions of the predicted networks with the sequential link thresholds for the ‘phosphorylation’ biomolecular network.} The panels with dotted lines (blue) signal that the degree distribution follows the power law. The marked panels were used in Fig. 3. The detailed network property is listed in Supplementary Table 3.
}
\label{figS7}
\end{figure*}

\begin{figure*}[p]
\begin{center}
\includegraphics[width=.9\linewidth]{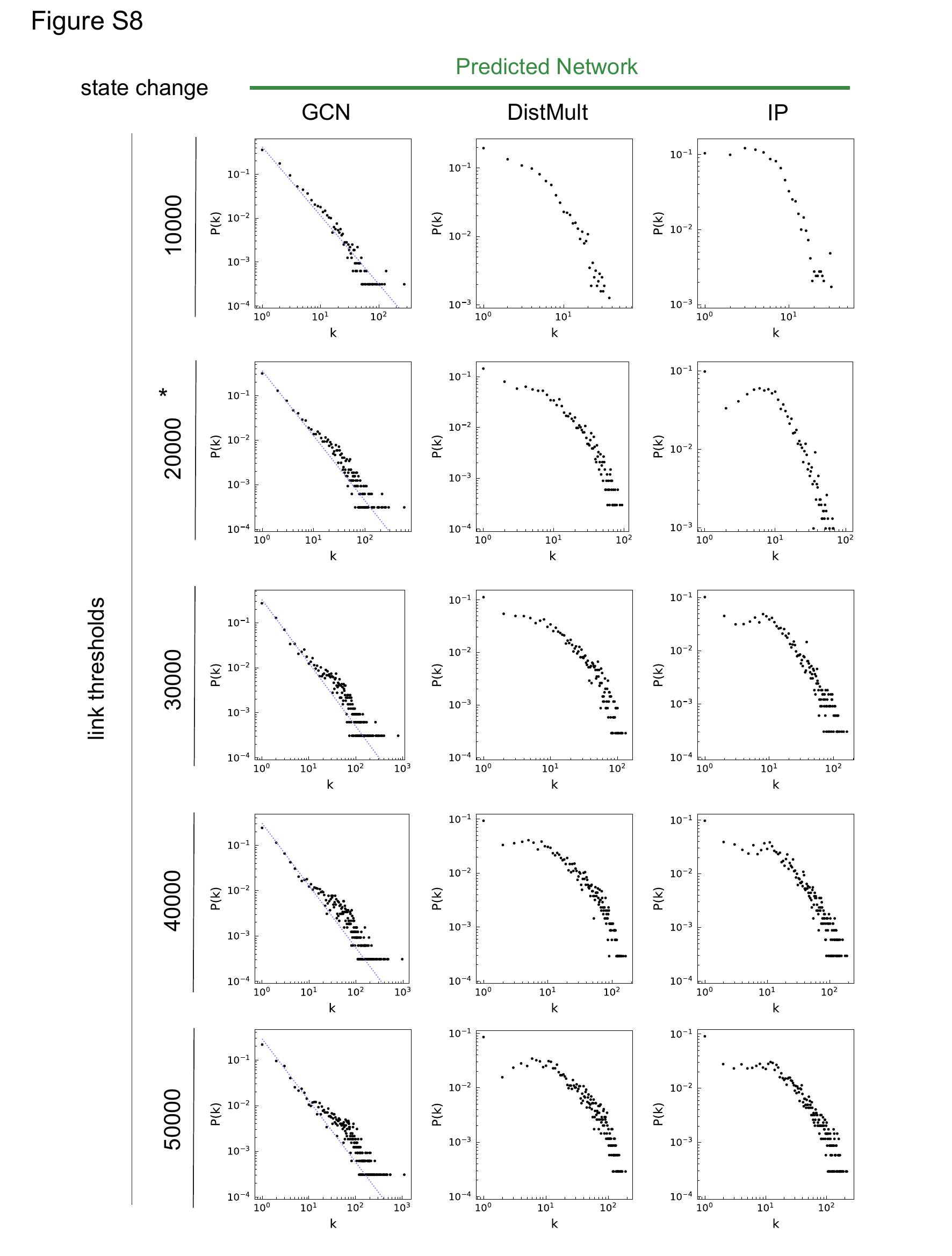}
\end{center}
\caption{\textbf{The degree distributions of the predicted networks with the sequential link thresholds for the ‘state change’ biomolecular network.} The panels with dotted lines (blue) signal that the degree distribution follows the power law. The marked panels were used in Fig. 3. The detailed network property is listed in Supplementary Table 3.
}
\label{figS8}
\end{figure*}

\begin{figure*}[p]
\begin{center}
\includegraphics[width=.9\linewidth]{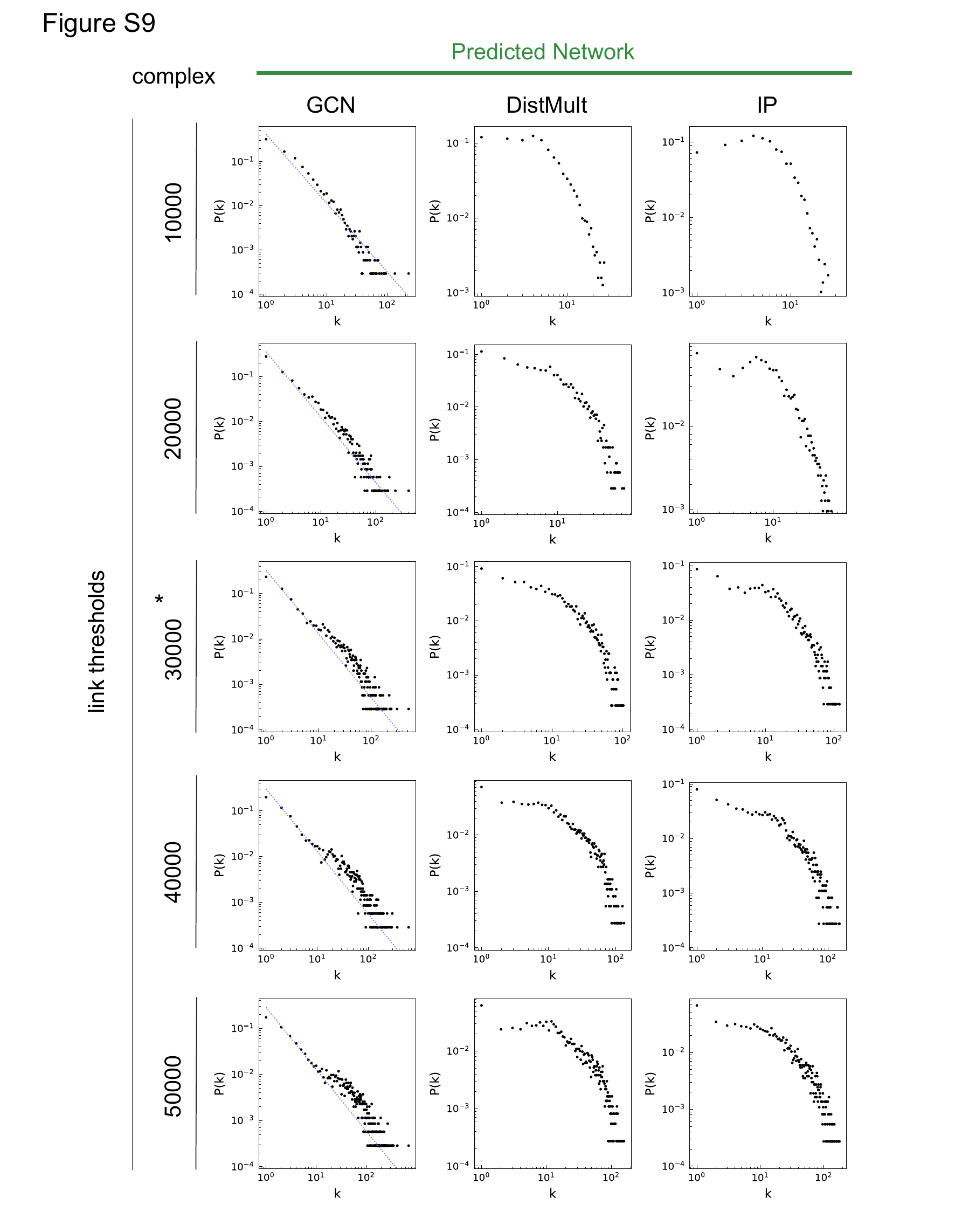}
\end{center}
\caption{\textbf{The degree distributions of the predicted networks with the sequential link thresholds for the ‘complex’ biomolecular network.} The panels with dotted lines (blue) signal that the degree distribution follows the power law. The marked panels were used in Fig. 3. The detailed network property is listed in Supplementary Table 3.
}
\label{figS9}
\end{figure*}

\begin{figure*}[p]
\begin{center}
\includegraphics[width=.9\linewidth]{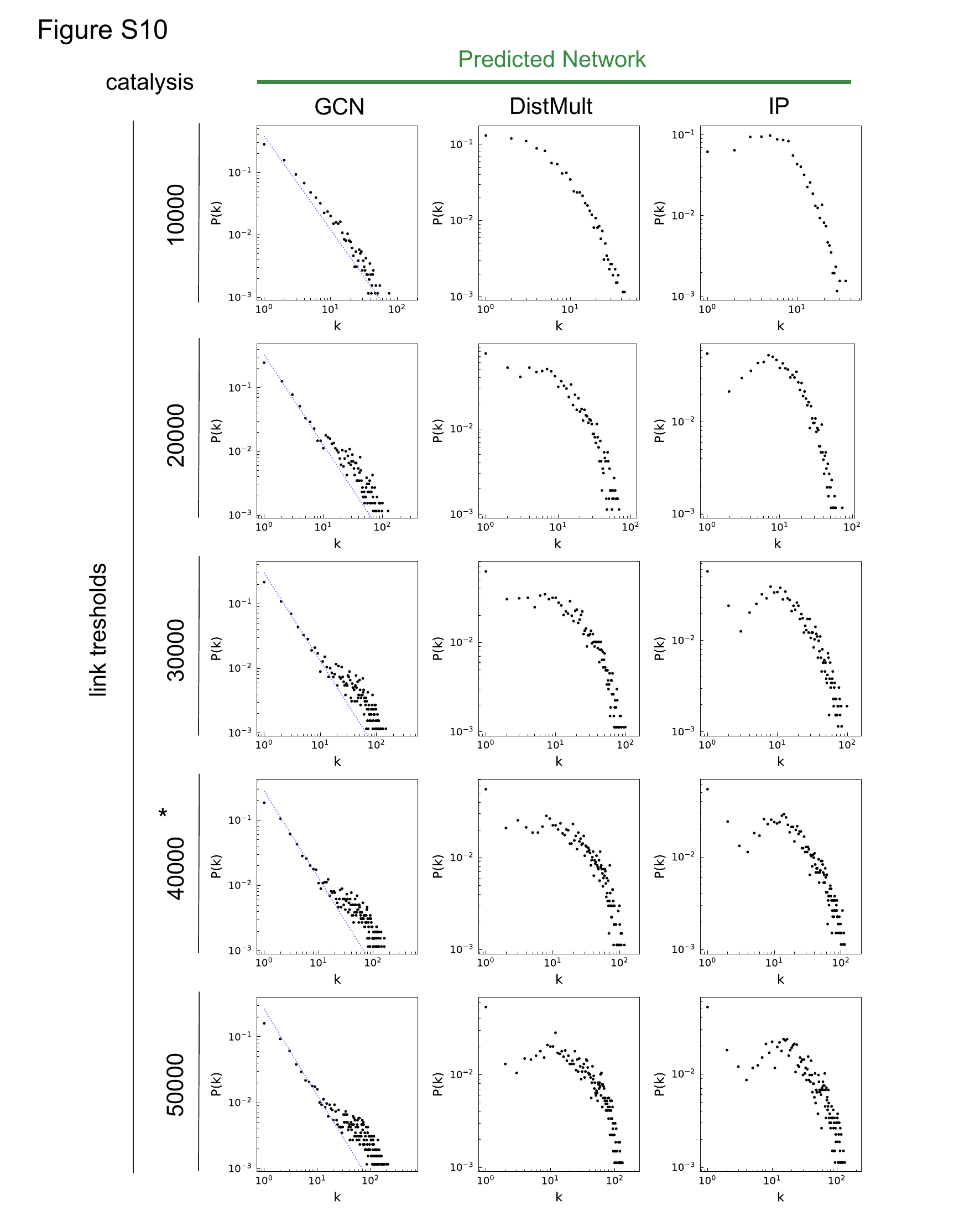}
\end{center}
\caption{\textbf{The degree distributions of the predicted networks with the sequential link thresholds for the ‘catalysis’ biomolecular network.} The panels with dotted lines (blue) signal that the degree distribution follows the power law. The marked panels were used in Fig. 3. The detailed network property is listed in Supplementary Table 3.
}
\label{figS10}
\end{figure*}

\begin{figure*}[p]
\begin{center}
\includegraphics[width=\linewidth]{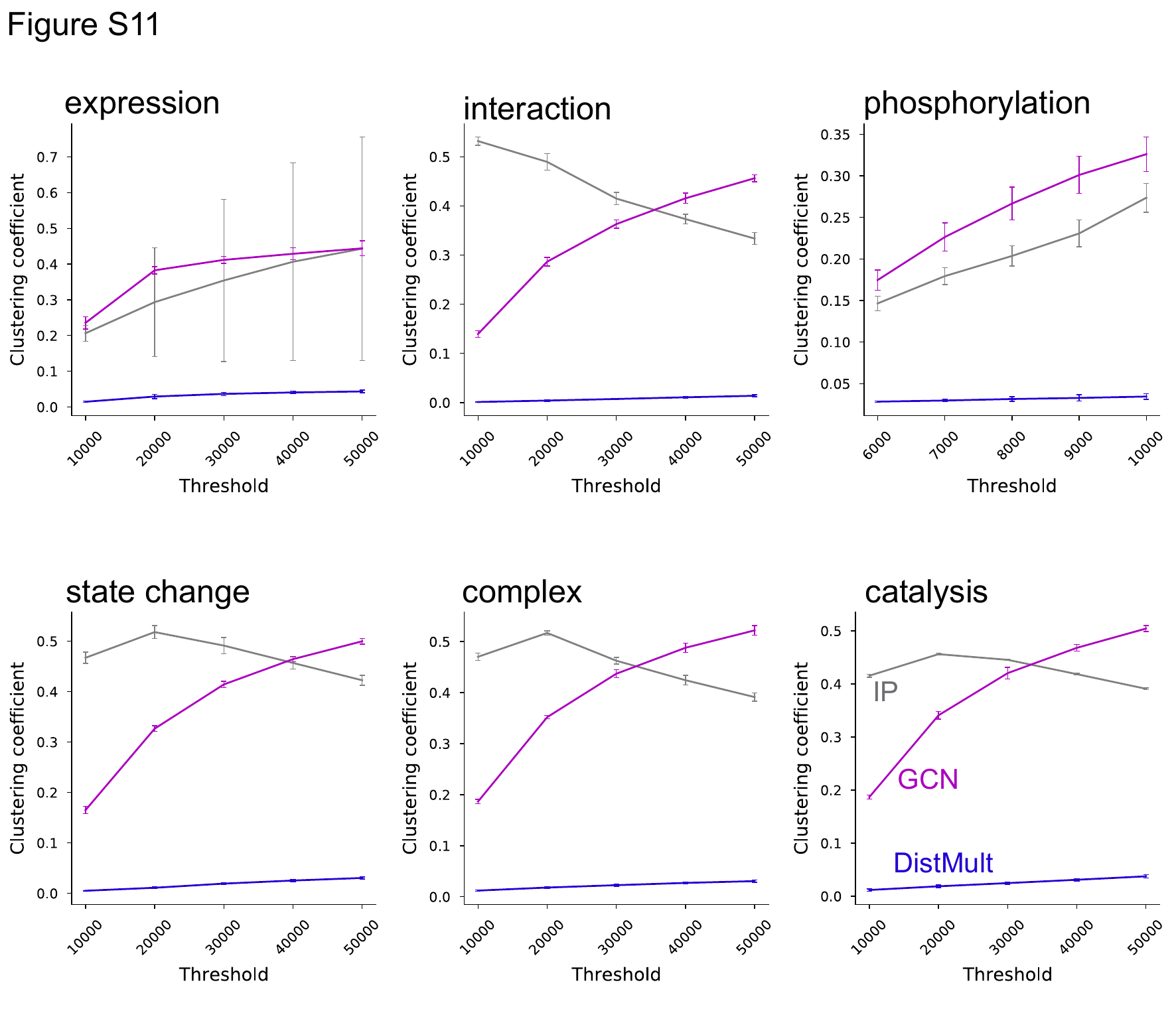}
\end{center}
\caption{\textbf{The transition of clustering coefficient for the six biomolecular networks.} The X axis represents the link threshold for the predicted networks. The detailed value is listed in Supplementary Table 3. Error bar: mean ± standard deviation (n = 3).
}
\label{figS11}
\end{figure*}

\begin{figure*}[p]
\begin{center}
\includegraphics[width=\linewidth]{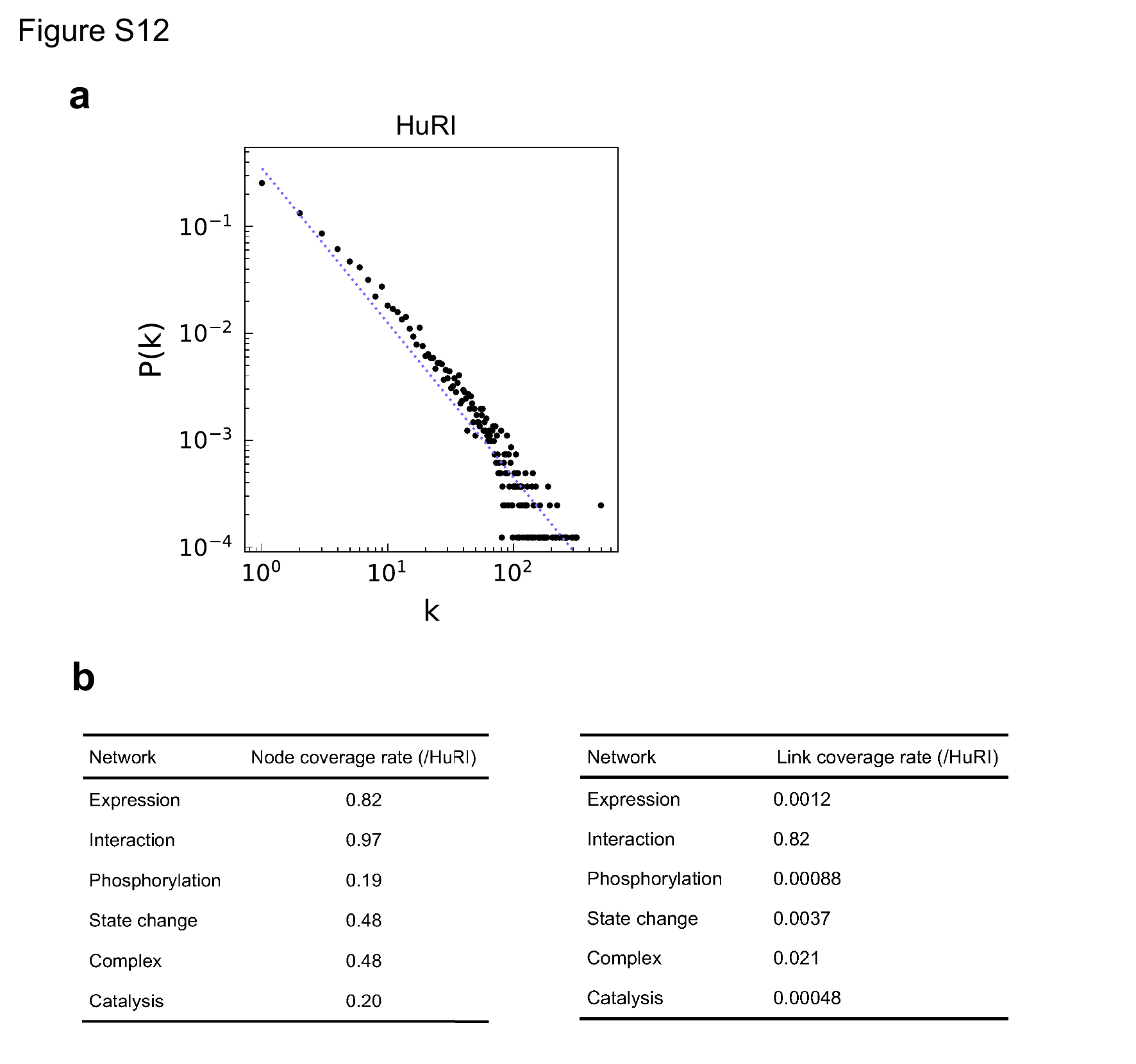}
\end{center}
\caption{\textbf{Characterization of the HuRI network.} (\textbf{a}) The degree distribution probability of the HuRI network. The X axis represents the degree k, and the Y axis the degree distribution probability P(k). The detailed network property is listed in Supplementary Table 1. (\textbf{b}) The coverage rate of nodes and links between the HuRI network and the six biomolecular networks.
}
\label{figS12}
\end{figure*}

\end{document}